# A PHENOMENOLOGICAL ENVIRONMENTAL TRAFFIC MODEL Part 1: Fuzzy Logic Validatation

S. Nicolosi[1], P. Ferrante[2], G. Scaccianovce[2], G. Peri[2], GF Rizzo[2]
[1]CNR/IBFM-lATO, Cefalù, Italy
[2]DEIM, Università degli studi di Palermo, Italy

**ABSTRACT**

In this first paper of a series of two we are going to show that, on the ground of few reasonable hypotheses, an equation linking the number of vehicle $N(\tau_i)$, that run from *(i-1)-th* to the *i-th* hour, to the ground-level hourly average measured values of an arbitrary pollutant concentration $[C(\tau_i ,z=0)\equiv C(\tau_i)]$ can be established. The traffic environmental model so generated is validated via a fuzzy logig approach. In a second paper we will valitadeted and calibrated it via a manual counting of traffic fluxes and its provisional capability will be tessted

## 1. INTRODUCTION

The research of a theoretical, model based, relationship linking emission sources and urban pollutant concentration as never received big attention. The most conventional approach is to look for an empirically-based general relationship linking concentration to the multitude of factors - emissions, meteorology, atmospheric chemistry, pollution cleaning – influencing air quality at various geographical scales (for a review [1]).
Despite this fact, a lot of works have been produced in order to, independently, model the pollutant turbulent dynamics and the traffic flow. The absence of a connection between the theories can be related to the big number of deficiencies showed by the theoretical traffic models [2-4] when applied to urban centres, dramatically affected by vehicular emissions. On the contrary the original approach proposed in this paper can be considered one of the first attempts to built a bridge between the theories.
In particular we are going to show that, on the ground of few reasonable hypotheses, an equation linking the number of vehicle $N(\tau_i)$, that run from *(i-1)-th* to the *i-th* hour, to the ground-level hourly average measured values of an arbitrary pollutant concentration $[C(\tau_i ,z=0)\equiv C(\tau_i)]$ can be established.
The main steps in achieving this goal are summarized in the figure 1.

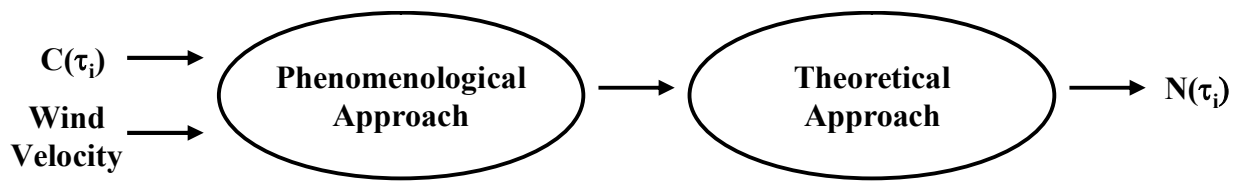

Figure 1

Referring to the situation of the running fleet and the measured pollutant concentrations concerning the Italian town of Palermo, a *data-deduced* traffic model is here derived. Unfortunately, we have not the possibility to validate this model by direct inspection of vehicular flow and, a part from the wind speed dependence of the local concentration, it is not simple to explain how the other microclimatic variables analytically influence its values. This is due, primarily, to the incompleteness and imprecision of our data, at this stage. In order to represent and manipulate incomplete databases, several extensions of the classical relational models have been proposed. In particular the fuzzy set theory, referring to the fuzzy logical proposed by Zadeh [5,6], provides a suitable mathematical framework for dealing with incomplete and imprecise information.

In this paper an adaptative fuzzy network is trained in order to simulate and to validate the proposed traffic models, as showed in the figure 2.

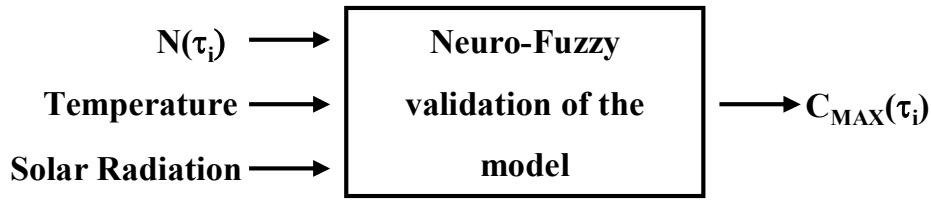

Figure 2

This traffic model has been calibrated by using the data of ground level pollutant concentration measured in the Italian town of Palermo by AMIA (Azienda Municipalizzata…….). In particular we considered the collected data at the receptor site named Piazza Castelnuovo. This choice is due to the fact that in this site the principal microclimatic variables (temperature, solar radiation, wind velocity and so on) are also monitored.

The measured data are referred to two years, namely 1997 and 2002, because they seem to be the less affected by systematic error in the monitoring process. These data are averaged out a period of one hour.

The measured values of concentration depend on a multitude of variables, that anyway can be usefully divided in three groups:
1. the variables always affecting the local concentrations in the same way, discarding from the particular instant of time in which the measurement is performed(the urban configuration) and whose influence will be entirely contained within some analytical constant parameter.
2. The set of local microclimatic conditions whose random variation hardly influences the concentration measurements.
3. The socio-demographic variables, whose influence can be related to the daily working-day traffic demand, in order to deduce quantitative schemes of the temporal distribution of the running vehicles number.

The main feature of these schemes is that the number and typology of vehicles running inside the urban network can be derived from socio-demographic data, as the annual census of population and housing, employment data and other relevant social input. The higher is the number of involved variables the higher is the reliability of the model. The interpretation frames in this way obtained can provide information about the average *per hour* number of the emission sources in an arbitrary working day.

Therefore, by accepting these models, we have implicitly assumed that every working day of the year is characterized by the same average *per hour* emissions. This hypothesis constitutes the basis for our data analysis. We will show that, on the ground of this assumption, a data-deduced traffic model can be built-up, so bypassing the problem of collecting socio-demographic data.

**Phenomenological Approach**

The first step in the construction of our traffic model is based on the observation of concentration data behaviour as functions of microclimatic variables.
Two main hypotheses have been assemed:
- hp 1 - The traffic demand does not depend on a particular working day, but only on the specific instant of time (of an arbitrary day) in which the measurement is performed.
- hp 2 – the measured values of concentration in the same interval of time of different working days can assume different values because of a variation in the microclimatic conditions and not because of a changing in the number of vehicles.

The first hypothesis implies that in the same time of different working day the pollutant emissions are constant. Moreover this constant emission supplies an upper limit for the ground level concentrations, since the other affecting-measurement variables can only cause a decrease of it.

For this reason we expect that the average per hour measured values of pollutant concentrations show the same behaviour, as a function of the time, in all the working days of the year.

This not means that the measured values are the same but that they have maxima or minima in the same interval of time and the evolution between two adjacent opposite limit is almost of the same kind.

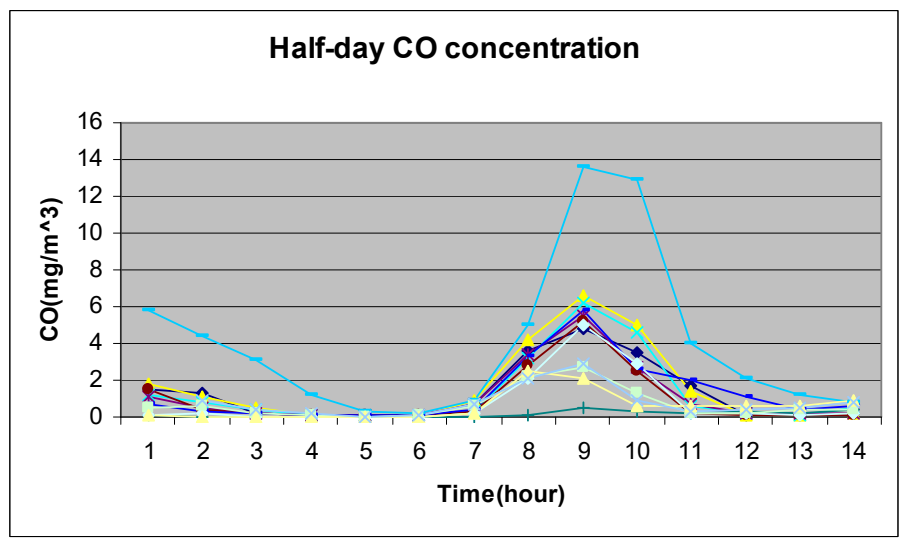

Figure 3. Tipical half day CO concentration

Figure 3 displays the half-day average per hour concentration of CO measured in the central square of Palermo, that is Piazza Castelnuovo, as a function of time in a few representative working day. The principal behaviour we can observe is that, from four o'clock to five o'clock in the morning, it takes its minimum value; then, it increases showing a maximum value at nine o'clock; subsequently, it decreases reaching a relative minimum at two o'clock p.m.

Let us suppose now that all the random variables, except the wind speed, are constant and equal to the values for which the actual ground level concentration is the maximum one. If we could perform a measurement of the concentrations, in the interval of time going from eight to nine o' clock a.m., under these ideal conditions, we would expect to find a set of points belonging to the curve that represents the functional dependence of the concentration from the wind speed. If, now, we think to leave the other parameters free to change, we would find in the same plane a cloud of points confusedly distributed inside a limited area, whose upper bound is given by the curve first identified.

This simple reasoning shows that we can obtain the analytical relationship linking concentrations and wind speeds under the ideal, previously identified, conditions, and that it has to be represented by the boundary line, providing the separation between the areas with and without points.

Further, if our arguments are corrects, changing the observation time or, equivalently, varying the number of emission sources, the cloud of points should be characterised by the same features with a boundary line that is lower than the previous one.

Figure 4 shows the overlapping of different series, corresponding to different intervals of time (from 3 to 9 a.m.), of the ground-level hourly average measured values of CO concentration $[C(\tau,z=0) \equiv C(\tau)]$ as a function of the hourly average wind speed measured values. The figure appears to be in accordance with our arguments and, besides, the overlapping of the different coloured series singles out the family of *iso-emissive* curves. From these curves it appears that, increasing the wind speed, the concentration decreases following an exponential wise law. This behaviour is reasonable enough: in fact, when the value of the wind speed is zero the concentration takes its maximum allowed value; when the speed goes to infinity, concentration goes to zero discarding all other parameters.

The figure 5 shows the same concentration data in a logarithm scale.

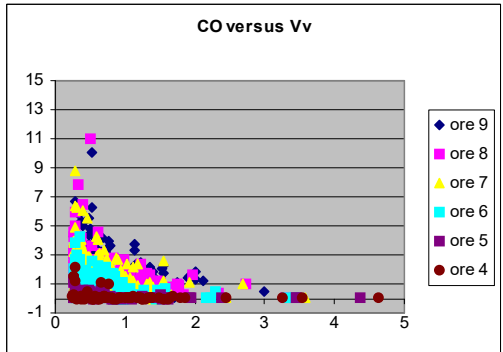 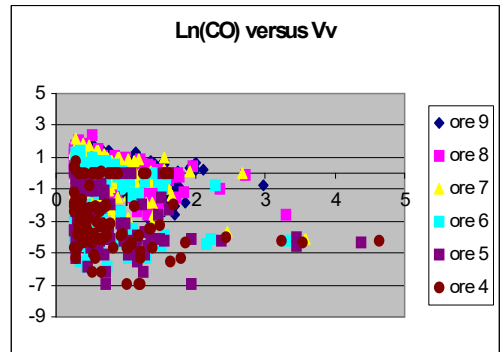

Figura 4           Figura 5

The isoemissive curves are straight lines of the same slope and different *intercepta*, whose equation is the following:

$$Ln(C(\tau_i)) = C_{MAX}(\tau_i) - kV \quad , \quad i=1,...,24 \quad (1)$$

or, equivalently,

$$C(\tau_i) = C_{MAX}(\tau_i)e^{-kV} \quad (2)$$

with k≈3/2 as deducible by a linear fitting of figure 5.

The times $\tau_i$ are discrete intervals of one hour amplitude, so that $\tau_i$ is the interval of time ranging from the hour *(i-1)-th* and the hour *i-th, i=1,...,24*. The constant *k* depends on the urban configuration, while $C_{MAX}(\tau_i)$, that is the hourly average value of concentration at the ground level corresponding to a windless condition, has to be related to the analyzed pollutant and to the number of vehicles running in the same interval of time.

**Theoretical Approach**
The theoretical approach can be summarized by te graph showed in the figure 6.

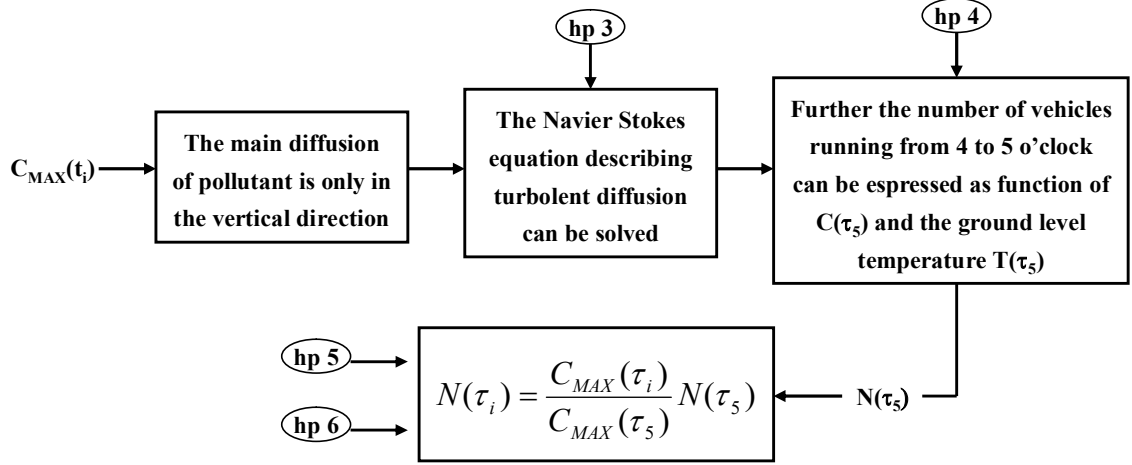

In order to find the analytical relationship between $C_{MAX}(\tau_i)$ and the number of emission sources, we can assume, without loosing generality, that the turbulent motion of pollutant is only along vertical direction. In fact, the advective ground-level motion of pollutant is phenomenological kept into account by the, previously derived, exponential law. Therefore our attention is, now, focused on the windless ground level concentration $C_{MAX}(\tau_i)$ of an arbitrary pollutant.
Moreover discarding the friction with the building wall, this turbulent motion causes the volume of the pollutants to increase without suffering of any deformation.
The diffusion along the vertical direction is a free expansion of an ideal gas under the influence of the gravitational field of the temperature gradient.
The complete model describing the dynamical behaviour of the gas incorporates as variables the density, the mass concentration the velocity and the temperature of the pollutant. The model is based on the assumption of local laws of balance of momentum, energy and mass resulting in the well known Navier Stokes and continuity equations.
These equations form asset of mutually interconnected, nonlinear, three-dimensional, time-dependent, partial differential equation.
The complexity of the equations has brought us to introduce the "night-time pollutant equilibrium hypothesis" by virtue of which the dynamical problem becomes a static exactly resoluble problem.
The hypothesis can be formulated as follow:
  hp 3    – from 4 o'clock to 5 o'clock in the morning the registered level of concentration are so much low that it is reasonable to assume that, in this interval of time the concentration may reach an equilibrium configuration, because of the low number of sources and the fastness of turbulent diffusion processes (whose characteristic time can be assumed much lower than one hour);
Under this hypothesis the total mass emitted from the vehicles running from four to five o'clock a.m. is [nostro lavoro]:

$$M_E(5) = \frac{AC_{MAX}(\tau_5)T(\tau_5)R}{mg}, \quad (3)$$

Were A is the area of Piazza Castelnuovo, $T(\tau_5)$ is the ground level average temperature in the previously selected interval of time, R is the constant of ideal gases, m is the molecular mass of the pollutant and g is gravitational acceleration of Earth.

If now we indicate with S the volumetric emission rate from pollution sources and introduce the following assumptions:

- hp 4     - with reference to a given year, it's possible to define the "yearly average vehicle" (YAV) [7] of a whole modality of transport, that is the amount of pollutants emitted by an average vehicle that is assumed as representative of whole running fleet of Palermo for an unitary length of travel;

We can write the total emitted pollutant mass in the time interval $\tau_i$ as:

$$M_E(\tau_i) = N(\tau_i) YAV <p>, \quad (4)$$

where $<p>$ is the average route of the $N(\tau_i)$ vehicles.

Inserting eq. (3) in the previous one we obtain a linear equation for the unknown quantity $N(\tau_5)$ that can be easily solved so obtaining:

$$N(\tau_5) = \frac{AC_{MAX}(\tau_5)T(\tau_5)R}{mg(YAV)<p>} \quad (5)$$

The last step of the theoretical approach is based on the further following assumption:

- hp 5     - it appears reasonable to assume that $C_{MAX}(\tau_i)$ is proportional to the global emission appeared in the *i-th* hour, reduced by an unknown scale factor $f(X(i),Y(i),...)$ depending on the local microclimatic variable $X(i),Y(i),...$, with $i=1,...,24$, that is:

$$C_{MAX}(\tau_i) = N(\tau_i)(YAV)f(X(i),Y(i),...), \quad 0 \leq f(X(i),Y(i),...) \leq 1 \quad (3)$$

- hp 6     - we assume that $X(i) \approx X(i-1)$, $Y(i) \approx Y(i-1)$,...then $f(X(i),Y(i),...) \equiv f(X(i-1),Y(i-1),...)$.

On the ground of these hypotheses it is possible to show that:

$$\frac{C_{MAX}(\tau_i)}{C_{MAX}(\tau_{i-1})} = \frac{N(\tau_i)}{N(\tau_{i-1})} \quad (6)$$

and that:

$$N(\tau_i) = \frac{C_{MAX}(\tau_i)}{C_{MAX}(\tau_5)} N(\tau_5). \quad (7)$$

A statistical analysis of the calculated $N(\tau_i)$ shows that the frequency of appearance of each $N(\tau_i)$ follows a Gaussian distribution. It is, indeed, reasonable to take the mean values of the calculated sets of vehicular number as the best estimation of the average

per hour number of vehicles and its standard deviation as the associated error. In this way we obtain:

$$< N(\tau_9) > = 16590 \pm 7760$$
$$< N(\tau_8) > = 13940 \pm 6170$$
$$< N(\tau_7) > = 6860 \pm 3150$$
$$< N(\tau_6) > = 2270 \pm 1420$$
$$< N(\tau_5) > = 1190 \pm 1050$$

The traffic model we have generated is entirely based on hypotheses that cannot be validated in a direct way. For this reason, in the next session we will try to develop arguments that, in some respect, give a justification of the truthfulness of our traffic model.

## 3. FUZZY LOGIC

The analysis developed inside the traffic model section, starting from theoretical assumptions, discards the complex dependence of the measured value of concentration from the other local microclimatic variables, only taking into account its relationships to the wind speed. In this way we have introduced a traffic model able to give information about the number of circulating vehicles, once the hypotheses to make it working are satisfied. Unfortunately we haven't the possibility to test our traffic model by a direct inspection of the traffic flow.

In general, the measured concentration will take a value belonging to the statistical mixture of all permitted values, although conditioned by the actual value of the random microclimatic conditions.

Therefore the possibility of singling out the analytical relationship linking the average of the *per hour* windless pollutant concentration to the local microclimatic conditions, as well as to the local vehicular fleet, is here supplied by means of a fuzzyfication of the equation (12):

$$C_{MAX}(\tau_i) = N(\tau_i)(YAV) f(X(i), Y(i),...) \quad (15)$$

This equation is one of the ground hypotheses on which the traffic model is built on.
The choice of the inputs $X(i), Y(i)...$ is based on a proper account of "*where the pollutant went*" and "*what happened to it*". Assuming that:
- the pollutant is released to the environment by the *turbulent dispersion* flow due to the atmospheric structure (vertical temperature distribution);
- the pollutant undergoes photochemical transformations;
- the pollutant level is increased by emissions from pollutant sources.

The steps involved in the fuzzy adaptive learning process are:
- Fuzzifier the set of data
- Training a fuzzy neural network;
- the assignment of the membership functions to the input variables;
- the assignment of coarse dates-deduced rules linking inputs and outputs;
- the comparison between the theoretical outputs and the measured ones;

- the minimization of errors by the back-propagation or hybrid algorithm resulting in a correction of the previous assignments.

In the contest of air pollutant dispersion in urban environment the input variables are represented by a few metheoclimatic conditions. After a several studies of data set, relative to 1997 and 2002 years, it is ensued that the main metheoclimatic variables for the input are the temperature and solar radiation. The last used input is the number of vehicles, previously calculated by phenomenological and theoretical approaches. The output data set is represented by the concentration of Cmax of CO and NOx previously calculated. The trend of these results that we expect is a shaded trend in function of increase of Temperature and solar radiation.

Before the fuzzifier of data set these are modified to discharge a few data which aren't representative of the phenomenon. As we can see in the next figure there are a few data with a strange value that could invalidate the result of fuzzyfication model.

The first step of the fuzzyfication process consists in the assignment of the training and checking data sets. The lack of an objective criterion in choosing these data sets and starting from the availability of the fuzzy network to represent the pollutant behaviour has induced us to make a mixing of these sets in order of avoiding a preferential role. The distribution of the training and checking data sets, as measured, is shown in the figure 7.

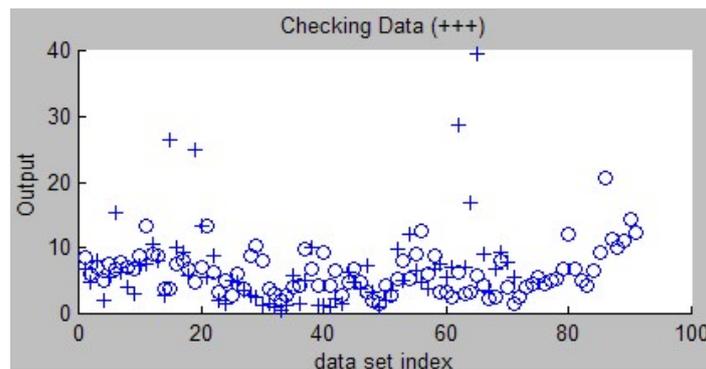

**Figure 7. The distribution of training and checking data sets.**

The assignment of the membership functions to the input variables and the definition of rules linking inputs and output has been made on the ground of the observation of the graphs depicting the pollutant as a function of temperature and solar radiation. There are two kind of membership functions, in this work we have used the Sugeno type. This type reproduce our output in the best way. As a matter of fact the outputs are represented by singletons, in particular the $C_{max}$ values of selected pollutant.

The comparison between the theoretical output and the measured ones has introduced an error whose minimization is provided by means of a hybrid algorithm (a combination of backpropagation and least-squares method) resulting in a correction of the previous assignments (membership functions and rules).

After about 120 epochs of training the neural fuzzy network has reached a steady value of the error (*e=0,07*) and has produced the distribution of training test data displayed in figure 8.

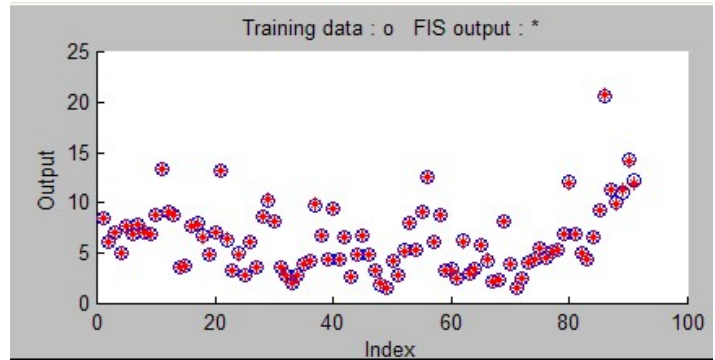
**Figure 8. Distribution of the trading test**

The small error, along with the full overlapping of the measured data with the theoretical output, ensures that the learning process has been correctly implemented. Moreover, the neural adaptive fuzzy interference system, whose principal structure is shown in the figure 9, has produced the analytical dependence of windless concentration as a function of the vehicular fleet showed in figure 10.

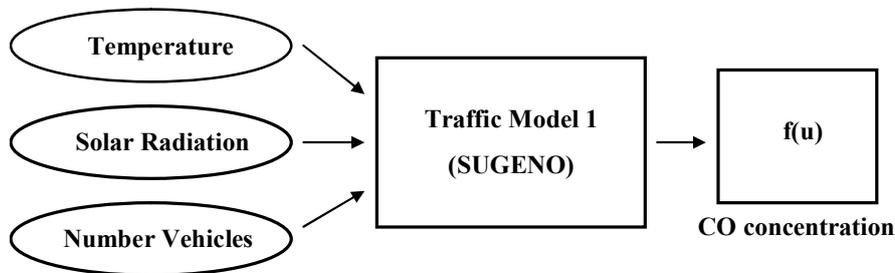
**Figure 9. Fuzzy inference system of traffic model**

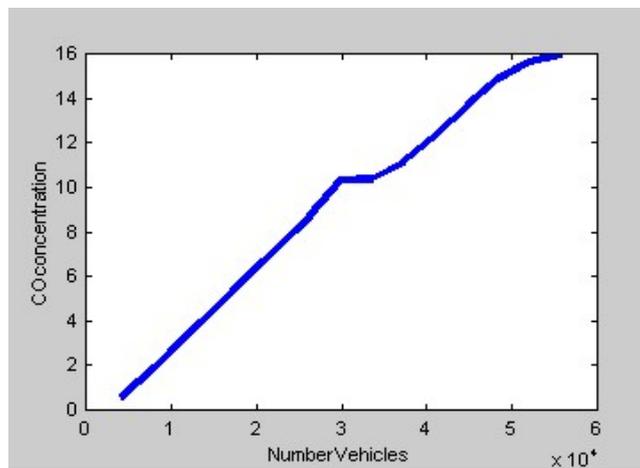
**Figure 10. The relationship between number of vehicles and CO concentrations**

The last figure predicts that the CO concentration increases as the number of running vehicles rises. A deviation from this behaviour occurs if this number is grater than 30.000 vehicles. In this regime the chaotic and dynamical diffusive phenomenon probably invalidates the ground hypotheses upon which the model has been derived.

## 5-CONCLUSIONS

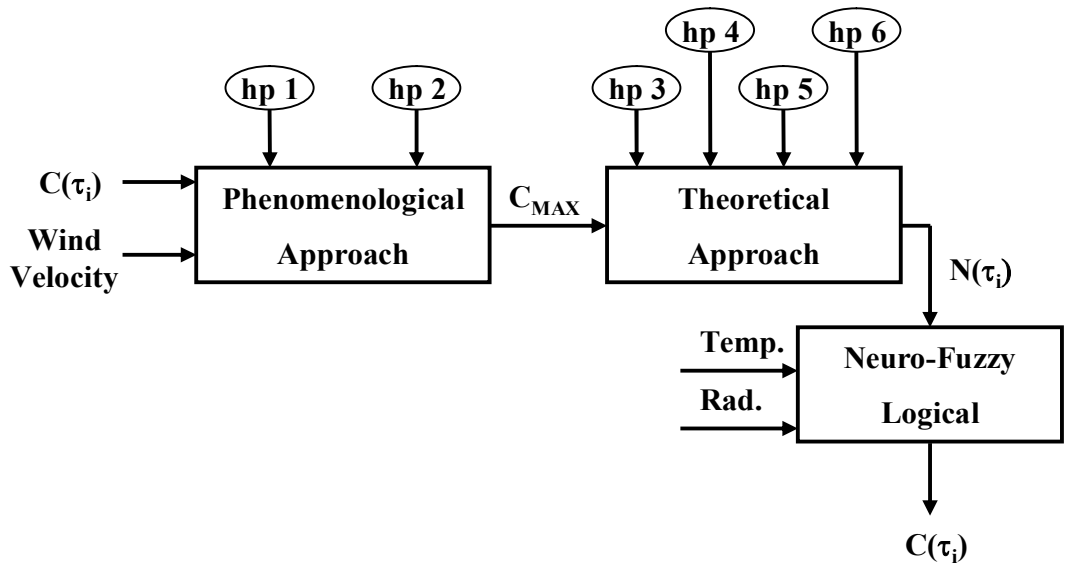

This work takes its place in the wide research activity concerning the air quality level in urban centres.

Its main result is the implementation of a traffic model able to relate the measured values of CO concentration to the running fleet of vehicles that, as it is well known, can be considered the main source of urban pollution.

The method here developed, although requiring further investigations, contains the novelty of deducing the average *per hour* number of vehicles on the basis of a few reasonable hypotheses.

The truthfulness of this hypotheses probably needs major speculative arguments, obviously. Anyway, the reasonability of the first outcome of the present *data-deduced* traffic model, clearly encourage to better investigate where the hypotheses could fall down, in the aim of producing a predictive tool to support local administrators in the decision-making processes.

The dispersion of the pollutants and the related measured concentration have a strong dependence from several parameters such as: the local microclimatic conditions, the vehicular fleet composition, the urban configuration and the socio-demographic variables, whose influence can be related to the daily working day traffic demand.